\documentclass[aps,preprint,nofootinbib,eqsecnum,superscriptaddress]{revtex4}
\pdfoutput=1
\usepackage[latin9]{inputenc}
\usepackage{float}
\usepackage{amsmath}
\usepackage{amssymb}
\usepackage{graphicx}
\usepackage{esint}
\usepackage{color}
\usepackage{float}
\usepackage{bm}
\usepackage{epstopdf}
\usepackage{latexsym}

\makeatletter
\@ifundefined{textcolor}{}
{
 \definecolor{BLACK}{gray}{0}
 \definecolor{WHITE}{gray}{1}
 \definecolor{RED}{rgb}{1,0,0}
 \definecolor{GREEN}{rgb}{0,1,0}
 \definecolor{BLUE}{rgb}{0,0,1}
 \definecolor{CYAN}{cmyk}{1,0,0,0}
 \definecolor{MAGENTA}{cmyk}{0,1,0,0}
 \definecolor{YELLOW}{cmyk}{0,0,1,0}
}
\@ifundefined{textcolor}{}{
 \definecolor{BLACK}{gray}{0}
 \definecolor{WHITE}{gray}{1}
 \definecolor{RED}{rgb}{1,0,0}
 \definecolor{GREEN}{rgb}{0,1,0}
 \definecolor{BLUE}{rgb}{0,0,1}
 \definecolor{CYAN}{cmyk}{1,0,0,0}
 \definecolor{MAGENTA}{cmyk}{0,1,0,0}
 \definecolor{YELLOW}{cmyk}{0,0,1,0}
}
\@ifundefined{textcolor}{}{
 \definecolor{BLACK}{gray}{0}
 \definecolor{WHITE}{gray}{1}
 \definecolor{RED}{rgb}{1,0,0}
 \definecolor{GREEN}{rgb}{0,1,0}
 \definecolor{BLUE}{rgb}{0,0,1}
 \definecolor{CYAN}{cmyk}{1,0,0,0}
 \definecolor{MAGENTA}{cmyk}{0,1,0,0}
 \definecolor{YELLOW}{cmyk}{0,0,1,0}
}
\def\be{\begin{equation}}
\def\ee{\end{equation}}
\def\bea{\begin{eqnarray}}
\def\eea{\end{eqnarray}}
\def\ba{\begin{array}}
\def\ea{\end{array}}

\makeatother

\begin{document}

\title{Post-transient relaxation in graphene after an intense laser pulse}
\author{Junhua Zhang}
\affiliation{Department of Physics, College of William and Mary, Williamsburg, Virginia
23187, USA}
\author{Tianqi Li}
\affiliation{Ames Laboratory and Department of Physics and Astronomy, Iowa State
University, Ames, IA 50011, USA}
\author{Jigang Wang}
\affiliation{Ames Laboratory and Department of Physics and Astronomy, Iowa State
University, Ames, IA 50011, USA}
\author{J\"{o}rg Schmalian}
\affiliation{Institute for Theory of Condensed Matter and Center for Functional
Nanostructures, Karlsruhe Institute of Technology, Karlsruhe 76128, Germany}
\date{\today\\
\vspace{0.6in} }

\begin{abstract}
High intensity laser pulses were recently shown to induce a population
inverted transient state in graphene {[}T. Li \emph{et al.} Phys. Rev. Lett. 
\textbf{108}, 167401 (2012){]}. Using a combination of hydrodynamic
arguments and a kinetic theory we determine the post-transient state
relaxation of hot, dense, population inverted electrons towards equilibrium.
The cooling rate and charge-imbalance relaxation rate are determined from
the Boltzmann-equation including electron-phonon scattering. We show that
the relaxation of the population inversion, driven by inter-band scattering
processes, is much slower than the relaxation of the electron temperature,
which is determined by intra-band scattering processes. This insight may be
of relevance for the application of graphene as an optical gain medium.
\end{abstract}

\maketitle

\section{Introduction}

\begin{figure}[tbp]
\hfill {}\includegraphics[scale=0.8]{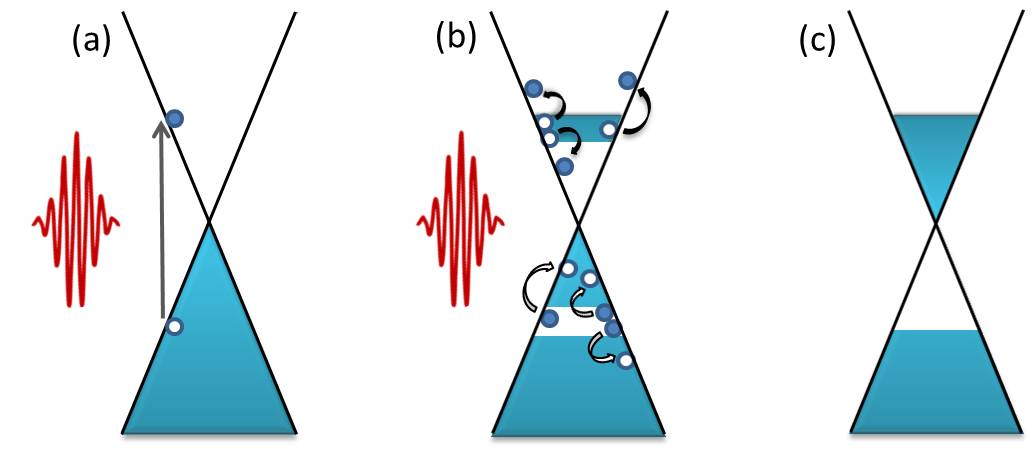}\hfill {}
\caption{Schematic demonstration of the formation of population inverted
electronic state right after an intense laser pulse. (a) Photoexcited
carriers generated by $\sim $10fs pump pulse; (b) The leading scattering
processes of photo-excited carriers taking place in several femtoseconds: $%
e+e\rightarrow e+e,\ h+h\rightarrow h+h,\ e+h\rightarrow e+h$ ($e$-electron
carriers and $h$-hole carriers), which quickly establish individual
thermalization in electron and hole carriers sharing a common electronic
temperature $T$ due to the electron-hole scattering events; (c) After the
internal thermalization, the photoexcited carriers form a population
inverted hourglass-like electronic state characterized by distinct chemical
potentials $\protect\mu _{+}$, $\protect\mu _{-}$, and a common electron
temperature $T$. }
\label{fig: demonstration}
\end{figure}

Recently, it was shown that photoinduced femtosecond nonlinear saturation,
transparency and stimulated infrared emission of extremely dense fermions in
graphene monolayers emerge\cite{Li2012}. A single laser pulse of $35$ fs
quasi-instantaneously builds up a broadband, inverted Dirac fermion
population, where optical gain emerges and manifests itself via a negative
optical conductivity, see Fig.\thinspace \ref{fig: demonstration}. 
Increasing the excitation from the linear to the highly nonlinear regime,
the photoexcited transient state evolves from a hot classical gas to a dense
quantum fluid. Such high-density population inversion at femtosecond time
scales has significant implications in advancing graphene based above
terahertz speed modulators, saturable absorbers, or an ultra-broadband gain
medium. These results emerge in a regime where the photoexcited carrier
density is much larger than the initial background carriers and one is no
longer in the linear power dependence of transient signals. \ An important
open question in this context is the origin of the comparatively stable
population inverted state despite the rapid thermalization. In Ref.\cite%
{Li2012} we argued, based on results obtained using perturbation theory with
respect to the electron-electron Coulomb interaction\cite%
{Fritz2008,Mueller2009,Foster2009}, that the transient state of dense Dirac
fermions is stabilized by the phase-space constraints of the relativistic
spectrum. The change in the optical conductivity as function of photoexcited
carriers was then very well described in terms of a nonequilibrium electron
distribution function. For this distribution function we assumed the
quasi-equilibrium form 
\begin{equation}
f_{\mathbf{k}\lambda }=\frac{1}{e^{\left( \varepsilon _{\mathbf{k}\lambda
}-\mu _{\lambda }\right) /\left( k_{B}T\right) }+1},  \label{distr}
\end{equation}%
characterized by the linear dispersion relation $\varepsilon _{\mathbf{k}%
\lambda }=\lambda v\hbar \left\vert \mathbf{k}\right\vert $, where $\lambda
=+\left( -\right) $ refers to the upper (lower) branch of the graphene
spectrum with velocity $v$. $T$ is the electron temperature and $\mu
_{\lambda }$ refer to the chemical potentials that are allowed to be
distinct for the upper and lower branch of the spectrum. The population
inversion is therefore characterized by $\delta \mu =(\mu _{+}-\mu _{-})/2$.
Results for the electron temperature and the chemical potentials were
determined from an analysis of the energy and charge balance of the system%
\cite{Li2012,Zhang2013}.

A natural question to ask is the nature of the post-transient relaxation
that gives rise to a relaxation of the laser induced population inversion
back to equilibrium. In addition to the electron-electron Coulomb
interaction, the post-transient regime is characterized by the coupling of
the electronic systems to the lattice, which is expected to lead to a
relaxation of the electronic energy and population inversion. The
investigation of this question is the subject of this manuscript. Our theory
is a generalization of the approach used in Refs.\cite%
{Bistrizter2009,DasSarma01,DasSarmaold} to the regime of population inverted
initial states. In Refs.\cite{Bistrizter2009,DasSarma01} the energy transfer
to phonons was analyzed as the dominant low-temperature cooling channel of
excited electrons in graphene without population inversion. Based on the
assumption of a rapid thermalization of the electron system the change in
the electron temperature 
\begin{equation}
\frac{\partial T\left(t\right)}{\partial t}=Q/C
\end{equation}
was determined from the electronic heat capacity $C$ and the cooling power $%
Q=\frac{\partial\mathcal{E}\left(t\right)}{\partial t}$. \ This approach is
justified in a hydrodynamic regime, where the characteristic change of the
temperature $\left\vert \frac{1}{T\left(t\right)}\frac{\partial
T\left(t\right)}{\partial t}\right\vert $ is much smaller than the
microscopic relaxation rate $\tau^{-1}$. The cooling power $Q$ was then
determined from an analysis of the Boltzmann equation\cite%
{Bistrizter2009,DasSarma01}.

Here we generalize this approach and include the corresponding change in the
chemical potentials $\mu_{\lambda}$ into account, i.e. we analyze the
distribution function Eq.\ref{distr} with time dependent electron
temperature and chemical potential: $T\rightarrow T\left(t\right)$ and $%
\mu_{\lambda}\rightarrow\mu_{\lambda}\left(t\right)$. This enables us to
monitor the temporal evolution of the nonequilibrium state that follows the
intense laser pulse and compare the dynamics of the electron heating and
population inversion $\delta\mu\left(t\right)$. In the first part of our
theory section we give a summary of the key hydrodynamic relations that
apply to our system. In a second step we give explicit results for the
cooling power and imbalance relaxation that are obtained from an analysis of
the corresponding kinetic equation.

\section{Theory}

\subsection{Hydrodynamic considerations}

We analyze the time evolution of the transient state that is characterized
by an effective electron temperature $T\left(t\right)$ and chemical
potentials $\mu_{\lambda}\left(t\right)$ following an intense laser pulse.
The latter induces electron heating and a population inversion, see Fig.\,%
\ref{fig: demonstration}. Within a hydrodynamic description, analogous to
Refs.\cite{Bistrizter2009,DasSarma01,DasSarmaold}, we use a quasistatic
description. To this end we analyze the internal energy $\mathcal{E}%
\left(T,\mu_{+},\mu_{-}\right)$ and the particle numbers $%
N_{\lambda}\left(T,\mu_{\lambda}\right)$ of the two branches of the graphene
spectrum: 
\begin{eqnarray}
d\mathcal{E} & = & C_{e}dT+\sum_{\lambda}\frac{\partial\mathcal{E}%
\left(T,\mu_{+},\mu_{-}\right)}{\partial\mu_{\lambda}}d\mu_{\lambda},  \notag
\\
dN_{\lambda} & = & \chi_{\lambda}dT+\kappa_{\lambda}d\mu_{\lambda}.
\label{11}
\end{eqnarray}
Here we made the assumption that the occupation of the $\lambda$-th branch
of the spectrum only depends on its own chemical potential $\mu_{\lambda}$,
and not on the chemical potential of the other branch. This assumption will
be justified later in explicit calculations of the involved kinetic
processes. In Eq.\ref{11} we used the heat capacity $C$, the
compressibilities $\kappa_{+}$ and $\kappa_{-}$\ of the upper and lower
Dirac cone, respectively, as well as the corresponding changes in the
occupations as function of temperature $\chi_{+}$ and $\chi_{-}$: 
\begin{eqnarray}
C\left(T,\mu_{+},\mu_{-}\right) & = & \frac{\partial\mathcal{E}%
\left(T,\mu_{+},\mu_{-}\right)}{\partial T},  \notag \\
\kappa_{\lambda}\left(T,\mu_{\lambda}\right) & = & \frac{\partial
N_{\lambda}\left(T,\mu_{\lambda}\right)}{\partial\mu_{\lambda}},  \notag \\
\chi_{\lambda}\left(T,\mu_{\lambda}\right) & = & \frac{\partial
N_{\lambda}\left(T,\mu_{\lambda}\right)}{\partial T}.
\end{eqnarray}
The change in energy as function of chemical potential $\partial\mathcal{E}%
\left(T,\mu_{+},\mu_{-}\right)/\partial\mu_{\lambda}=\left.\partial\mathcal{E%
}/\partial\mu_{\lambda}\right\vert _{T}$, can be expressed in terms of $%
\chi_{\lambda}$. The corresponding Maxwell relation is given as $%
\left.\partial\mathcal{E}/\partial\mu_{\lambda}\right\vert
_{T}=-N_{\lambda}+T\chi_{\lambda}$.

For quasiparticles with distribution function Eq.\ref{distr} \ these
response functions are given as 
\begin{eqnarray}
C_{e} & = & \frac{N_{d}}{T^{2}}\int_{\mathbf{k},\lambda}\left(\varepsilon_{%
\mathbf{k}\lambda}-\mu_{\lambda}\right)^{2}f_{\mathbf{k}\lambda}\left(1-f_{%
\mathbf{k}\lambda}\right),  \notag \\
\kappa_{\lambda} & = & \frac{N_{d}}{T}\int_{\mathbf{k}}f_{\mathbf{k}%
\lambda}\left(1-f_{\mathbf{k}\lambda}\right),  \notag \\
\chi_{\lambda} & = & \frac{N_{d}}{T^{2}}\int_{\mathbf{k}}\left(\varepsilon_{%
\mathbf{k}\lambda}-\mu_{\lambda}\right)f_{\mathbf{k}\lambda}\left(1-f_{%
\mathbf{k}\lambda}\right),  \label{heat cap etc}
\end{eqnarray}
where $N_{d}=4$ refers to the valley and spin degeneracy of graphene. We use
the notation $\int_{\mathbf{k},\lambda}\cdots=\int\frac{d^{2}k}{%
\left(2\pi\right)^{2}}\sum_{\lambda}\cdots$ and $\int_{\mathbf{k}}\cdots=\int%
\frac{d^{2}k}{\left(2\pi\right)^{2}}\cdots$, and set $k_{B}=1$ in the
expressions. The time evolution of these quasistatic states is then
determined by 
\begin{eqnarray}
\frac{\partial\mathcal{E}}{\partial t} & = & C\frac{\partial T}{\partial t}%
+\sum_{\lambda}\left(-N_{\lambda}+T\chi_{\lambda}\right)\frac{%
\partial\mu_{\lambda}}{\partial t},  \notag \\
\frac{\partial N_{\lambda}}{\partial t} & = & \chi_{\lambda}\frac{\partial T%
}{\partial t}+\kappa_{\lambda}\frac{\partial\mu_{\lambda}}{\partial t}.
\label{hydrochange}
\end{eqnarray}
Since the evolution should be done under the condition of fixed total charge 
$\frac{\partial N}{\partial t}=\sum_{\lambda}\frac{\partial N_{\lambda}}{%
\partial t}=0$, we can express the time dependence of the mean chemical
potential $\overline{\mu}$ as $\frac{\partial\overline{\mu}}{\partial t}=-%
\frac{\delta\kappa}{\overline{\kappa}}\frac{\partial\delta\mu}{\partial t}-%
\frac{\overline{\chi}}{\overline{\kappa}}\frac{\partial T_{e}}{\partial t}$,
where $\overline{\chi}=\sum_{\lambda}\chi_{\lambda}$, $\overline{\kappa}%
=\sum_{\lambda}\kappa_{\lambda}$, $\delta\chi=\sum_{\lambda}\lambda\chi_{%
\lambda}$, $\delta\kappa=\sum_{\lambda}\lambda\kappa_{\lambda}$, and $%
\mu_{\lambda}=\overline{\mu}+\lambda\delta\mu$. If we introduce the charge
imbalance $\Delta=\sum_{\lambda}\lambda N_{\lambda}$, we finally obtain for
the change in energy $\mathcal{E}$ and $\Delta$:

\begin{eqnarray}
\frac{\partial\mathcal{E}}{\partial t} & = & C_{e}\frac{\partial T}{\partial
t}+\sum_{\lambda}\left(-N_{\lambda}+T\chi_{\lambda}\right)\frac{%
\partial\mu_{\lambda}}{\partial t},  \notag \\
\frac{\partial\Delta}{\partial t} & = & \left(\delta\chi-\frac{\delta\kappa%
\overline{\chi}}{\overline{\kappa}}\right)\frac{\partial T}{\partial t}%
+\left(\overline{\kappa}-\frac{\delta\kappa^{2}}{\overline{\kappa}}\right)%
\frac{\partial\delta\mu}{\partial t}.
\end{eqnarray}
In order to have explicit expressions for $\frac{\partial\mathcal{E}}{%
\partial t}$ and $\frac{\partial\Delta}{\partial t}$ we next resort to a
kinetic theory.

\subsection{Analysis of the kinetic equation}

Next we analyze the Boltzmann equation that leads to an energy and imbalance
relaxation due to the coupling to optical and acoustic phonons. The time
evolution of the transient state with effective temperature and chemical
potentials, caused by the coupling to phonons, is then determined by the
distribution function $f_{\mathbf{k}\lambda}\left(t\right)$. The time
evolution of the distribution function Eq.\ref{distr} follows from the
Boltzmann equation 
\begin{equation}
\frac{\partial f_{\mathbf{k}\lambda}\left(t\right)}{\partial t}=I_{\mathbf{k}%
\lambda}\left(t\right),
\end{equation}
with collision term: 
\begin{eqnarray*}
I_{\mathbf{k}\lambda} & = & -\frac{2\pi}{\hbar}\int_{\mathbf{k}%
^{\prime},\lambda^{\prime}}\left\vert g_{\mathbf{k,k}^{\prime}}^{\lambda%
\lambda^{\prime},a}\right\vert ^{2}\left[f_{\mathbf{k}\lambda}\left(1-f_{%
\mathbf{k}^{\prime},\lambda^{\prime}}\right)+\left(f_{\mathbf{k}\lambda}-f_{%
\mathbf{k}^{\prime},\lambda^{\prime}}\right)n_{\mathbf{q},a}\right]%
\delta\left(\varepsilon_{\mathbf{k}\lambda}-\omega_{\mathbf{k-k}^{\prime}%
\mathbf{,}a}-\varepsilon_{\mathbf{k}^{\prime}\mathbf{,}\lambda^{\prime}}%
\right) \\
& & +\frac{2\pi}{\hbar}\int_{\mathbf{k}^{\prime},\lambda^{\prime}}\left\vert
g_{\mathbf{k,k}^{\prime}}^{\lambda\lambda^{\prime},a}\right\vert ^{2}\ \left[%
f_{\mathbf{k}^{\prime},\lambda^{\prime}}\left(1-f_{\mathbf{k}%
\lambda}\right)-\left(f_{\mathbf{k}\lambda}-f_{\mathbf{k}^{\prime},\lambda^{%
\prime}}\right)n_{\mathbf{q},a}\right]\ \delta\left(\varepsilon_{\mathbf{k}%
\lambda}+\omega_{\mathbf{k-k}^{\prime}\mathbf{,}a}-\varepsilon_{\mathbf{k}%
^{\prime}\mathbf{,}\lambda^{\prime}}\right),
\end{eqnarray*}
where $g_{\mathbf{k,k}^{\prime}}^{\lambda\lambda^{\prime},a}$ represent the
coupling coefficients of the $\lambda$-band electrons of wavevector $\mathbf{%
k}$ with the $a$-mode phonons to yield a $\lambda^{\prime }$-band electrons
of $\mathbf{k}^{\prime }$. Similar to the hydrodynamic analysis presented
above, we analyze the total energy and the particle numbers of the two
branches: 
\begin{eqnarray}
\mathcal{E}\left(t\right) & = & N_{d}\int_{\mathbf{k},\lambda}\left(%
\varepsilon_{\mathbf{k}\lambda}-\mu_{\lambda}\left(t\right)\right)f_{\mathbf{%
k}\lambda}\left(t\right),  \notag \\
N_{\lambda}\left(t\right) & = & N_{d}\int_{\mathbf{k}}f_{\mathbf{k}%
\lambda}\left(t\right).
\end{eqnarray}
From the Boltzmann equation follows 
\begin{eqnarray}
\frac{\partial\mathcal{E}}{\partial t} & = & N_{d}\int_{\mathbf{k}%
,\lambda}\left(\varepsilon_{\mathbf{k}\lambda}-\mu_{\lambda}\right)I_{%
\mathbf{k}\lambda}-N_{d}\int_{\mathbf{k},\lambda}f_{\mathbf{k}\lambda}\frac{%
\partial\mu_{\lambda}}{\partial t},  \notag \\
\frac{\partial\Delta}{\partial t} & = & N_{d}\int_{\mathbf{k}%
,\lambda}\lambda I_{\mathbf{k}\lambda}.  \label{BBEE}
\end{eqnarray}
The changes in the energy and particle number are now determined by the
collision integral of electron-phonon scattering.

We can also make contact between this approach and the hydrodynamic
considerations presented earlier. The quasi-equilibrium form, Eq.\ref{distr}
implies that 
\begin{equation}
\frac{\partial f_{\mathbf{k}\lambda}}{\partial t}=f_{\mathbf{k}%
\lambda}\left(1-f_{\mathbf{k}\lambda}\right)\left[\frac{\left(\varepsilon_{%
\mathbf{k}\lambda}-\mu_{\lambda}\right)}{T_{e}^{2}}\frac{\partial T}{%
\partial t}+\frac{1}{T}\frac{\partial\mu_{\lambda}}{\partial t}\right],
\end{equation}
which yields 
\begin{eqnarray}
\frac{\partial\mathcal{E}}{\partial t} & = & -N_{d}\int_{\mathbf{k},\lambda}%
\frac{\partial\mu_{\lambda}}{\partial t}f_{\mathbf{k}\lambda}
\label{qeqevol} \\
& & +N_{d}\int_{\mathbf{k},\lambda}f_{\mathbf{k}\lambda}\left(1-f_{\mathbf{k}%
\lambda}\right)\left[\frac{\left(\varepsilon_{\mathbf{k}\lambda}-\mu_{%
\lambda}\right)^{2}}{T^{2}}\frac{dT}{dt}+\frac{\varepsilon_{\mathbf{k}%
\lambda}-\mu_{\lambda}}{T}\frac{d\mu_{\lambda}}{dt}\right]  \notag \\
\frac{\partial\Delta}{\partial t} & = & N_{d}\int_{\mathbf{k,}\lambda}f_{%
\mathbf{k}\lambda}\left(1-f_{\mathbf{k}\lambda}\right)\lambda\left[\frac{%
\left(\varepsilon_{\mathbf{k}\lambda}-\mu_{\lambda}\right)}{T^{2}}\frac{%
\partial T}{\partial t}+\frac{1}{T}\frac{\partial\mu_{\lambda}}{\partial t}%
\right]  \notag
\end{eqnarray}
This yields our earlier result Eq.\ref{hydrochange}, including the
expressions Eq.\ref{heat cap etc} for the electronic heat capacity, $C$, the
compressibility $\kappa_{\lambda}=\partial N_{\lambda}/\partial\mu_{\lambda}$
and the change of the particle numbers with temperature $\chi_{\lambda}=%
\partial N_{\lambda}/\partial T$. This analysis also justifies our earlier
assumption that the particle number of one branch does not explicitly depend
on the other chemical potential. Of course, this is a direct consequence of
the form Eq.\ref{distr} of the distribution function.

We finally obtain a coupled set of equations that determines the time
evolution of the population inversion $\delta\mu$ and electron temperature $T
$: 
\begin{eqnarray}
\left(\frac{C}{T}-\frac{\overline{\chi}^{2}}{\overline{\kappa}}\right)\frac{%
\partial T}{\partial t}+\left(\delta\chi-\frac{\delta\kappa\overline{\chi}}{%
\overline{\kappa}}\right)\frac{\partial\delta\mu}{\partial t} & = &
N_{d}\int_{\mathbf{k},\lambda}\frac{\varepsilon_{\mathbf{k}%
\lambda}-\mu_{\lambda}}{T}I_{\mathbf{k}\lambda}  \notag \\
\left(\delta\chi-\frac{\delta\kappa\overline{\chi}}{\overline{\kappa}}\right)%
\frac{\partial T}{\partial t}+\left(\overline{\kappa}-\frac{\delta\kappa^{2}%
}{\overline{\kappa}}\right)\frac{\partial\delta\mu}{\partial t} & = &
N_{d}\int_{\mathbf{k},\lambda}\lambda I_{\mathbf{k}\lambda}
\end{eqnarray}
In what follows we analyze this set of equations numerically. We present our
data as function of the typical time scale for $\tau=4\pi^{2}\hbar^{3}v^{2}/%
\left(a_{0}^{2}g^{2}\delta\mu\left(0\right)\right)$ with electron phonon
coupling constant $g$, that characterizes intra-band scattering processes in
the regime where the initial population inversion $\delta\mu\left(0\right)$
is large compared to the phonon frequency. Here, $a_{0}=1.42\times10^{-10}%
\mathrm{m}$ is the C-C bond length.

\section{\protect\bigskip{} Results}

Our analysis of the time-evolution of the electron temperature $T(t)$ and
the population inversion $\delta\mu\left(t\right)$ is performed for graphene
at the neutrality point. We determine the initial electron temperature $%
T\left(t=0\right)$ and chemical potentials $\mu_{\lambda}\left(t=0\right)$
from the charge and energy balance right after the pulse; for details see 
\cite{Li2012,Zhang2013}. Our time $t=0$ point refers to the beginning of the
post transient evolution, i.e. about $100$fs after the initial laser pulse
that caused the population inverted state in the first place. In addition to
the value of the Fermi velocity, and the equilibrium chemical potentials and
temperature, the parameters of the theory are the pump-laser frequency $%
\hbar\omega_{pump}=1.55$ eV and the number of photoexcited carriers $n_{%
\text{ex}}(t=0)$ that is determined from the pump-fluence of the laser. Here
we use $n_{\text{ex}}(t=0)=4\times10^{13}$cm$^{-2}$. \ We obtain $%
T\left(0\right)=3336.4$ K and $\delta\mu\left(0\right)=\mu_{+}\left(0%
\right)=-\mu_{-}\left(0\right)=0.549761$eV. Finally to describe the
relaxation due to optical phonons we use $\hbar\omega_{o}\approx196$ meV as
optical phonon energy. For the lattice temperature we assume that it stays
constant around room temperature $T_{L}=300\mathrm{K}$, i.e. we assume that
the heat is quickly transferred to the bulk of the substrate. This aspect
should be more subtle in case of suspended graphene. For simplicity we
assume that the frequency is momentum and phonon-branch independent. The
electron-optical phonon coupling constant $g^{2}\equiv\left\langle \left(g_{%
\mathbf{k},\mathbf{k}^{\prime}}^{\lambda\lambda^{\prime},o}\right)^{2}\right%
\rangle $ takes a typical value $g=$ $0.315\,$eV\cite{Butscher07}. Since we
present our results as function of $t/\tau$, only the overall time scales
are determined by the value of $g$. Using $\delta\mu(0)=0.55$ eV we obtain $%
\tau\simeq10.21$ps. The occurrence of the various temporal regimes that
follow from our theory are not affected by the value of $g$.

\begin{figure}[tbp]
\hfill {}\includegraphics[scale=0.85]{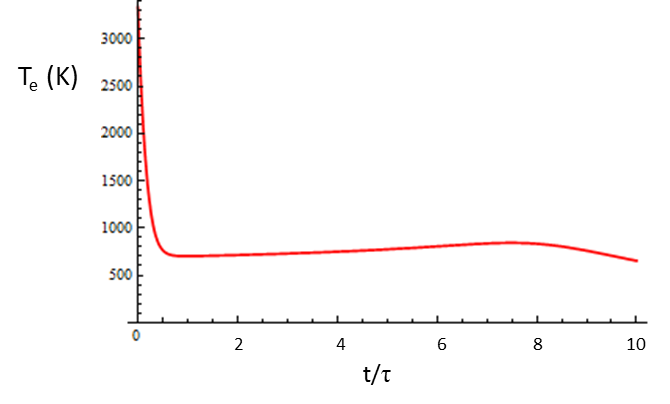}\hfill {}
\caption{Time evolution of the electron temperature $T(t)$ for a hot, dense,
population-inverted electron gas in graphene induced by an intense laser
pulse at $t<0$. The initial temperature of this system, $T\left( 0\right)
=3336.4$ K, follows a sharp drop within a short time period $t<\protect\tau %
/2$ mainly due to intra-band transitions, then reaches a plateau for a
relatively longer time period $\protect\tau /2<t<10\protect\tau $ before
further lowering down to the equilibrium temperature. Here $\protect\tau =4%
\protect\pi ^{2}\hbar ^{3}v^{2}/\left( a_{0}^{2}g^{2}\protect\delta \protect%
\mu \left( 0\right) \right) \simeq 10.21$ ps.}
\label{fig: Te_time}
\end{figure}

\begin{figure}[tbp]
\hfill {}\includegraphics[scale=0.85]{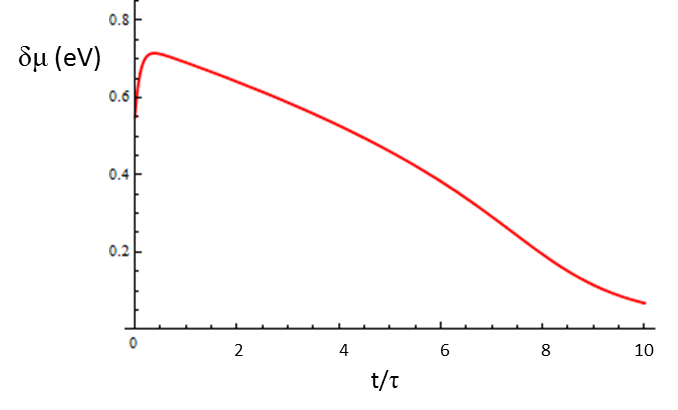}\hfill {}
\caption{Time evolution of the population inversion $\protect\delta \protect%
\mu (t)$ for a hot, dense, population-inverted electron gas in graphene at
the charge neutrality point where $\protect\delta \protect\mu (t)=\protect%
\mu _{+}(t)=-\protect\mu _{-}(t)$. The initial population inversion of this
system, $\protect\delta \protect\mu \left( 0\right) =0.549761$eV, follows a
small jump within the short time period $t<\protect\tau /2$ due to the
establishment of the sharp Fermi-distribution edge that is associated with
cooling, then decreases gradually as particle-hole recombination processes
through inter-band transitions become dominate for the longer time period$%
\protect\tau /2<t<10\protect\tau $. Here $\protect\tau \simeq 10.21$ ps.}
\label{fig: mu_time}
\end{figure}

\begin{figure}[tbp]
\hfill {}\includegraphics[scale=0.85]{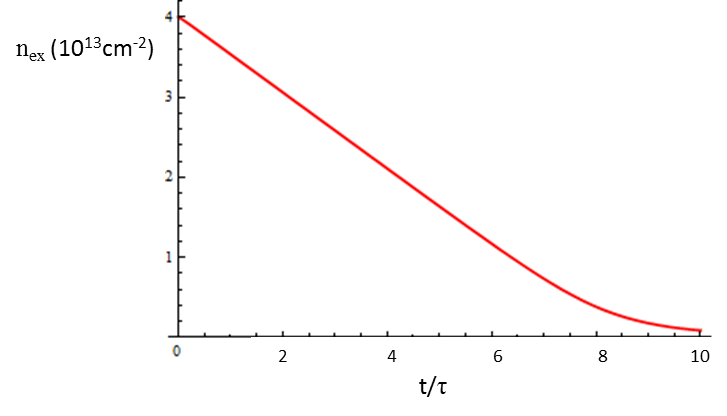}\hfill {}
\caption{Time evolution of the carrier density $n_{\mathrm{ex}}(t)$ for a
hot, dense, population-inverted electron gas in graphene. The photo-excited
carrier density takes an initial value at $n_{\text{ex}}(t=0)=4\times 10^{13}
$cm$^{-2}$, then drops gradually as a result of the inter-band carrier
scattering processes through coupling to optical phonons. Here $\protect\tau %
\simeq 10.21$ ps.}
\label{fig: Nex_time}
\end{figure}

In Fig.\,\ref{fig: Te_time}-\ref{fig: Nex_time} we show our results for the
changes of the electron temperature $T(t)$, the population inversion $%
\delta\mu(t)$, and the density $n_{\mathrm{ex}}(t)$ of photoexcited carriers
as function of time. The density of photoexcited carriers 
\begin{equation}
n_{\mathrm{ex}}=\frac{1}{2}\left(N_{+}-N_{+}^{0}-\left(N_{-}-N_{-}^{0}%
\right)\right)
\end{equation}
results from the calculated chemical potentials. Here $N_{\lambda}^{0}$ are
the particle densities of the two branches of the spectrum before the pulse.

\bigskip{}

Three distinct time regimes during the relaxation of a initially hot dense
state, with $k_{B}T(0)>\hbar\omega_{o}>k_{B}T_{L}$, are indicated from the
numerical result: At short time scale, $t/\tau<1/2$, the electron
temperature drops rapidly, while $\delta\mu$ rises due to the establishment
of the sharp Fermi-distribution edge that is associated with cooling. While $%
n_{\text{ex}}$, decreasing slowly and linearly, is not sensitive to the
rapid temperature drop and changes similar to the population inversion. The
fast initial cooling is mainly due to carrier intra-band transitions by
emitting optical phonons. This regime, dominated by carrier cooling process,
characterizes the energy relaxation of the system. Next we identify an
intermediate regime, $1/2<t/\tau<10$. In this regime we observe a plateau
for the evolution of the electron temperature, while $\delta\mu\left(t\right)
$ and $n_{\text{ex}}\left(t\right)$ decrease gradually. In this regime, the
carrier cooling driven by intra-band transitions is less efficient while a
relaxation dominated by inter-band transitions becomes dominant. This is the
regime where the population inversion is being destroyed. Particle-hole
recombination processes, characterize the population imbalance relaxation.
Finally, for longer times, $t/\tau>10-20$, the relaxation of population
inverted configurations is essentially completed and inter-band transitions
from the upper to the lower branch becomes less significant. What is left is
a slow cooling by the inefficient intra-band transition as $%
k_{B}T(t)\ll\hbar\omega_{o}$. In this regime, the coupling to acoustic
phonons, ignored in our treatment, should come into play and eventually
become the most dominant relaxation process for the terminal relaxation
towards equilibrium.

\section{Conclusions}

In conclusion, we investigated post-transient state relaxation of hot,
dense, population inverted electrons in graphene that emerged as the result
of an intense laser pulse, as shown in Ref.\cite{Li2012}. Using a
combination of hydrodynamic arguments and a kinetic theory we determined the
cooling rate and charge-imbalance relaxation rate. The latter are determined
from an analysis of the Boltzmann-equation where we included the scattering
between electrons and optical phonons. We demonstrated that the relaxation
of the electron temperature, driven by intra-band scattering processes, is
much more rapid than the relaxation of the population inversion, which is
determined by inter-band scattering processes. Thus, the relaxation of the
population inversion is significantly slower that the timescales responsible
for the energy transfers between the hot electron gas and the lattice. This
insight may be of relevance for the application of graphene as an optical
gain medium.

\section{\protect\bigskip{} Acknowledgment}

We thank Myron Hupalo and Michael Tringides for discussions. J.Z.
acknowledges support by the Jeffress Memorial Trust, Grant No. J-1033. J.S.
thanks the DFG Center for Functional Nanostructures. Work at Ames Laboratory
was partially supported by the U.S. Department of Energy, Office of Basic
Energy Science, Division of Materials Sciences and Engineering (Ames
Laboratory is operated for the U.S. Department of Energy by Iowa State
University under Contract No. DE-AC02-07CH11358).

\end{document}